\def\gsim{\;\lower4pt\hbox{${\buildrel\displaystyle >\over\sim}$}\,}
\def\lsim{\;\lower4pt\hbox{${\buildrel\displaystyle <\over\sim}$}\,}

\def\mag{{\bf B~}}
\def\avemag{$\langle{\bf B}\rangle$}
\def\FLASH{{\sc flash}}
\def\PARAMESH{{\sc paramesh}}

\documentclass{aa}  
\usepackage{graphicx}
\usepackage{natbib}
\usepackage[english]{babel}
\usepackage{txfonts}
\newcommand\rs[1]{_\mathrm{#1}}
\begin{document}
   \title{On the origin of asymmetries in bilateral supernova remnants}

   \author{S. Orlando\inst{1,2},
           F. Bocchino\inst{1,2},
           F. Reale\inst{3,1,2},
           G. Peres\inst{3,1,2}
          \and
           O. Petruk\inst{4,5}
          }

   \offprints{S. Orlando,\\ e-mail: orlando@astropa.inaf.it}

   \institute{INAF - Osservatorio Astronomico di Palermo ``G.S.
              Vaiana'', Piazza del Parlamento 1, 90134 Palermo, Italy
         \and
              Consorzio COMETA, via Santa Sofia 64, 95123 Catania, Italy
         \and
              Dip. di Scienze Fisiche \& Astronomiche, Univ. di
              Palermo, Piazza del Parlamento 1, 90134 Palermo,
              Italy
         \and 
              Institute for Applied Problems in Mechanics and
              Mathematics, Naukova St. 3-b Lviv 79060, Ukraine
         \and
              Astronomical Observatory, National University, Kyryla and
              Methodia St. 8 Lviv 79008, Ukraine
             }

   \date{Received \quad\quad\quad ; accepted \quad\quad\quad }

   \authorrunning{S. Orlando et al.}
   \titlerunning{On the origin of asymmetries in bilateral supernova remnants}

 
  \abstract
   {}
   {We investigate whether the morphology of bilateral supernova remnants
    (BSNRs) observed in the radio band is determined mainly either by
    a non-uniform interstellar medium (ISM) or by a non-uniform ambient
    magnetic field.}
   {We perform 3-D MHD simulations of a spherical SNR shock
    propagating through a magnetized ISM. Two cases of shock propagation
    are considered: 1) through a gradient of ambient density with a
    uniform ambient magnetic field; 2) through a homogeneous medium with
    a gradient of ambient magnetic field strength. From the simulations,
    we synthesize the synchrotron radio emission, making different
    assumptions about the details of acceleration and injection
    of relativistic electrons.}
   {We find that asymmetric BSNRs are produced if the line-of-sight is
   not aligned with the gradient of ambient plasma density or with the
   gradient of ambient magnetic field strength. We derive useful
   parameters to quantify the degree of asymmetry of the remnants that
   may provide a powerful diagnostic of the microphysics of strong shock
   waves through the comparison between models and observations.}
   {BSNRs with two radio limbs of different brightness can be explained
   if a gradient of ambient density or, most likely, of ambient magnetic
   field strength is perpendicular to the radio limbs. BSNRs with
   converging similar radio arcs can be explained if the gradient runs
   between the two arcs.}


   \maketitle
%
\section{Introduction}
\label{intro}

It is widely accepted that the structure and the chemical abundances
of the interstellar medium (ISM) are strongly influenced by supernova
(SN) explosions and by their remnants (SNRs). However, the details
of the interaction between SNR shock fronts and ISM depend, in principle,
on many factors, among which the multiple-phase structure of the medium,
its density and temperature, the intensity and direction of the ambient
magnetic fields. These factors are not easily determined and this
somewhat hampers our detailed understanding of the complex ISM.

The bilateral supernova remnants (BSNRs, \citealt{1998ApJ...493..781G};
also called "barrel-shaped," \citealt{1987A&A...183..118K}, or "bipolar",
\citealt{1990ApJ...357..591F}) are considered a benchmark for the
study of large scale SNR-ISM interactions, since no small scale effect
like encounters with ISM clouds seems to be relevant. The BSNRs are
characterized by two opposed radio-bright limbs separated by a region
of low surface brightness. In general, the remnants appear asymmetric,
distorted and elongated with respect to the shape and surface brightness
of the two opposed limbs. In most (but not all) of the BSNRs the symmetry
axis is parallel to the galactic plane, and this has been interpreted as
a difficulty for ``intrinsic'' models, e.g. models based on SN jets,
rather than for ``extrinsic'' models, e.g. models based on properties
of the surrounding galactic medium (\citealt{1998ApJ...493..781G}).

In spite of the interest around BSNRs, a satisfactory and complete model
which explains the observed morphology and the origin of the asymmetries
does not exist. The galactic medium is supposed to be stratified along the
lines of constant galactic latitude, and characterized by a large-scale
ambient magnetic field with field lines probably mostly aligned with
the galactic plane. The magnetic field plays a three-fold role: first,
a magnetic tension and a gradient of the magnetic field strength is
present where the field is perpendicular to the shock velocity leading
to a compression of the plasma; second, cosmic ray acceleration is
most rapid where the field lines are perpendicular to the shock speed
(\citealt{1987ApJ...313..842J}, \citealt{1988MNRAS.233..257O});
third, the electron injection could be favored where the magnetic
field is parallel to the shock speed (\citealt{1995ApJ...453..873E}).
\cite{1998ApJ...493..781G} notes that magnetic models (i.e. those
considering uniform ISM and ordered magnetic field) cannot explain the
asymmetric morphology of most BSNRs, and invokes a dynamical model
based on pre-existing ISM inhomogeneities, e.g. large-scale density
gradients, tunnels, cavities. Unfortunately, the predictions of these
ad-hoc models have consisted so far of a qualitative estimate of the BSNRs
morphology, with no real estimates of the ISM density interacting with the
shock. Moreover, most likely also non-uniform ambient magnetic fields may
cause asymmetries in BSNRs, without the need to assume ad-hoc density ISM
structures. Two main aspects of the nature of BSNRs, therefore, remain
unexplored: how and under which physical conditions do the asymmetries
originate in BSNRs? What is more effective in determining the morphology
and the asymmetries of this class of SNRs, the ambient magnetic field
or the non-uniform ISM?

Answering such questions at an adequate level requires detailed
physical modeling, high-level numerical implementations and extensive
simulations. Our purpose here is to investigate whether the morphology
of BSNR observed in the radio band could be mainly determined by the
propagation of the shock through a non-uniform ISM or, rather, across a
non-uniform ambient magnetic field. To this end, we model the propagation
of a shock generated by an SN explosion in the magnetized non-uniform ISM
with detailed numerical MHD simulations, considering two complementary
cases of shock propagation: 1) through a gradient of ambient density with
a uniform ambient magnetic field; 2) through a homogeneous isothermal
medium with a gradient of ambient magnetic field strength.

In Sect. \ref{sec2} we describe the MHD model, the numerical setup, and
the synthesis of synchrotron emission; in Sect. \ref{sec3} we analyze
the effects the environment has on the radio emission of the remnant;
finally in Sect. \ref{sec4} and \ref{sec5} we discuss the results and
draw our conclusions.

\section{Model}
\label{sec2}

\subsection{Magnetohydrodynamic modeling}

We model the propagation of an SN shock front through a magnetized ambient
medium. The model includes no radiative cooling, no thermal conduction, no
eventual magnetic field amplification and no effects on shock dynamics
due to back-reaction of accelerated cosmic rays. The shock propagation
is modeled by solving numerically the time-dependent ideal MHD equations
of mass, momentum, and energy conservation in a 3-D cartesian coordinate
system $(x,y,z)$:

\begin{equation}
\frac{\partial \rho}{\partial t} + \nabla \cdot (\rho \mbox{\bf u}) =
0~,
\end{equation}

\begin{equation}
\frac{\partial \rho \mbox{\bf u}}{\partial t} + \nabla \cdot (\rho
\mbox{\bf uu}-\mbox{\bf BB}) + \nabla P_* = 0~,
\end{equation}

\begin{equation}
\frac{\partial \rho E}{\partial t} +\nabla\cdot [\mbox{\bf u}(\rho
E+P_*) -\mbox{\bf B}(\mbox{\bf u}\cdot \mbox{\bf B})] = 0~,
\end{equation}

\begin{equation}
\frac{\partial \mbox{\bf B}}{\partial t} +\nabla \cdot({\bf uB}-{\bf
Bu}) = 0~,
\end{equation}

\noindent
where

\[
P_* = P + \frac{B^2}{2}~,~~~~~~~~~~~~~
E = \epsilon +\frac{1}{2} |\mbox{\bf u}|^2+\frac{1}{2}\frac{|\mbox{\bf
B}|^2}{\rho}~,
\]

\noindent
are the total pressure (thermal pressure, $P$, and magnetic pressure)
and the total gas energy (internal energy, $\epsilon$, kinetic energy,
and magnetic energy) respectively, $t$ is the time, $\rho = \mu m_H
n_{\rm H}$ is the mass density, $\mu = 1.3$ is the mean atomic mass
(assuming cosmic abundances), $m_H$ is the mass of the hydrogen atom,
$n_{\rm H}$ is the hydrogen number density, {\bf u} is the gas velocity,
$T$ is the temperature, and \mag is the magnetic field.  We use the
ideal gas law, $P=(\gamma-1) \rho \epsilon$, where $\gamma=5/3$
is the adiabatic index. The simulations are performed using the \FLASH\
code (\citealt{for00}), an adaptive mesh refinement multiphysics code
for astrophysical plasmas.

As initial conditions, we adopted the model profiles of
\citet{1999ApJS..120..299T}, assuming a spherical remnant with radius
$r\rs{0snr} = 4$~pc and with total energy $E_0 = 1.5\times 10^{51}$~erg,
originating from a progenitor star with mass of $15~M_{\rm sun}$, and
propagating through an unperturbed magnetohydrostatic medium. The
initial total energy is partitioned so that 1/4 of the SN energy is
contained in thermal energy, and the other 3/4 in kinetic energy. The
explosion is at the center $(x,y,z) = (0,0,0)$ of the computational
domain which extends between $-30$ and 30~pc in all directions. At the
coarsest resolution, the adaptive mesh algorithm used in the \FLASH\ code
(\PARAMESH; \citealt{mom00}) uniformly covers the 3-D computational
domain with a mesh of $8^3$ blocks, each with $8^3$ cells. We allow
for 3 levels of refinement, with resolution increasing twice at each
refinement level. The refinement criterion adopted (\citealt{loehner})
follows the changes in density and temperature. This grid configuration
yields an effective resolution of $\approx 0.1$ pc at the finest level,
corresponding to an equivalent uniform mesh of $512^3$ grid points.
We assume zero-gradient conditions at all boundaries.

We follow the expansion of the remnant for 22 kyrs, considering two sets
of simulations: 1) through a gradient of ambient density with a uniform
ambient magnetic field; or 2) through a homogeneous isothermal medium
with a gradient of ambient magnetic field strength. Table \ref{tab1}
summarizes the physical parameters characterizing the simulations.

\begin{table}
\caption{Relevant initial parameters of the simulations: $n_0$ and
$n_{i}$ are particle number densities of the stratified unperturbed
ISM (see text), $h$ is the density scale length, and $(x,y,z)$ are the
coordinates of the magnetic dipole moment. The ambient medium is either
uniform or with an exponential density stratification along the $x$
or the $z$ direction ($x-$strat. and $z-$strat., respectively); the
ambient magnetic field is uniform or dipolar with the dipole
oriented along the $x$ axis and located at $(x,y,z)$.}
\label{tab1}
\begin{tabular}{lcccccc}
\hline
\hline
 &  ISM & $n_0$ & $n_{i}$ & $h$ & \mag & $(x,y,z)$ \\
 &      & cm$^{-3}$ & cm$^{-3}$ & pc &    & pc  \\
\hline
GZ1 & $z-$strat. & 0.05 & 0.2 & 25 & uniform & - \\
GZ2 & $z-$strat. & 0.05 & 0.2 & 10 & uniform & - \\
GX1 & $x-$strat. & 0.05 & 0.2 & 25 & uniform & - \\
GX2 & $x-$strat. & 0.05 & 0.2 & 10 & uniform & - \\
DZ1 & uniform    & 0.25 & -   & -  & $z-$strat. & $(0,0,-100)$ \\
DZ2 & uniform    & 0.25 & -   & -  & $z-$strat. & $(0,0,-50)$ \\
DX1 & uniform    & 0.25 & -   & -  & $x-$strat. & $(-100,0,0)$ \\
DX2 & uniform    & 0.25 & -   & -  & $x-$strat. & $(-50,0,0)$ \\
\hline
\hline
\end{tabular}
\end{table}

In the first set of simulations, the ambient magnetic field
is assumed uniform with strength $\mag = 1~\mu$G and oriented
parallel to the $x$ axis. The ambient medium is modeled with an
exponential density stratification along the $x$ or the $z$ direction
(i.e. parallel or perpendicular to the \mag field) of the form: $n(\xi)
= n_0+n_i\exp(-\xi/h)$ (where $\xi$ is, respectively, $x$ or $z$)
with $n_0=0.05$ cm$^{-3}$ and $n_i = 0.2$ cm$^{-3}$, and where $h$
(set either to 25 pc or to 10 pc) is the density scale length. This
configuration has been used by e.g. \cite{1999A&A...344..295H} to
describe the SNR expansion in an environment with a molecular cloud.
Our choice leads to a density variation of a factor $\sim 6$ or $\sim
60$, respectively, along the $x$ or the $z$ direction over the spatial
domain considered (60 pc in total). The temperature of the unperturbed
ISM is $T=10^4$ K at $\xi=0$ and is determined by pressure balance
elsewhere. The adopted values of $T=10^4$ K, $n = 0.25$ cm$^{-3}$
and $B=1~\mu$G at $\xi=0$, outside the remnant, lead to $\beta \sim 17$
(where $\beta = P/(B^2/8\pi)$ is the ratio of thermal to magnetic
pressure) a typical order of magnitude of $\beta$ in the diffuse regions
of the ISM (\citealt{2004RvMP...76..125M}).

In the second set of simulations, the unperturbed ambient medium is
uniform with temperature $T=10^4$ K and particle number density $n = 0.25$
cm$^{-3}$. The ambient magnetic field, \mag, is assumed to be dipolar.
This idealized situation is adopted here mainly to ensure magnetostaticity
of the non-uniform field. The dipole is oriented parallel to the $x$
axis and located on the $z$ axis ($x=y=0$) either at $z=-100$ pc or at
$z=-50$ pc; alternatively the dipole is located on the $x$ axis ($y=z=0$)
either at $x=-100$ pc or at $x=-50$ pc (as shown in Fig.~\ref{fig1}). In
both configurations, the field strength varies by a factor $\sim 6$
($z$ or $x = -100$ pc) or $\sim 60$ ($z$ or $x = -50$ pc) over 60 pc:
in the first case in the direction perpendicular to the average ambient
field {\avemag}, whereas in the second case parallel to {\avemag}. In
all the cases, the initial magnetic field strength is set to $B =
1~\mu$G at the center of the SN explosion ($x=y=z=0$).

Note that the transition time from adiabatic to radiative phase for
a SNR is (e.g. \citealt{1998ApJ...500..342B, petruk05})

\begin{equation}
t\rs{tr} = 2.84\times
10^4\;E\rs{51}^{4/17}\;n\rs{ism}^{-9/17}~\mbox{yr}~,
\label{trans_time}
\end{equation}

\noindent
where $E\rs{51}=E\rs{0}/(10^{51}\ \mbox{erg})$ and $n\rs{ism}$ is the
particle number density of the ISM. In our set of simulations, runs GZ2
and GX2 present the lowest values of the transition time, namely
$t\rs{tr}\approx 25$ kyr. Since we follow the expansion of the remnant
for 22 kyrs, our modeled SNRs are in the adiabatic phase.

\subsection{Nonthermal electrons and synchrotron emission}
\label{e-acc}

We synthesize the radio emission from the remnant, assuming that it is
only due to synchrotron radiation from relativistic electrons distributed
with a power law spectrum $N(E)=KE^{-\zeta}$, where $E$ is the electron
energy, $N(E)$ is the number of electrons per unit volume with arbitrary
directions of motion and with energies in the interval $[E, E+dE]$,
$K$ is the normalization of the electron distribution, and $\zeta$
is the power law index. Following \cite{1965ARA&A...3..297G}, the radio
emissivity can be expressed as:

\begin{figure}[!t]
  \centering \includegraphics[width=8.5cm]{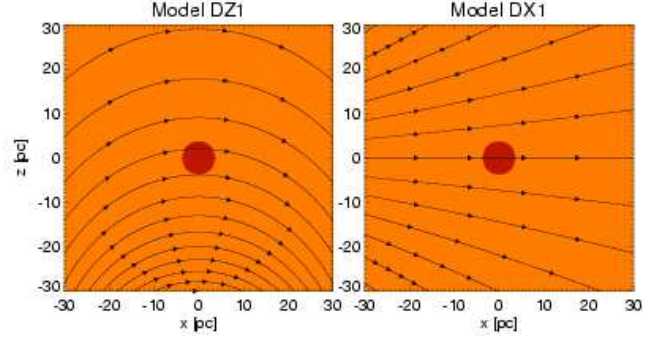}
  \caption{2-D sections in the $(x,z)$ plane of the initial mass density
           distribution and initial configuration of the unperturbed
           dipolar ambient magnetic field in two cases: dipole moment
           located on the $z$ axis (DZ1, left panel), or on the $x$ axis
           (DX1, right panel). The initial remnant is at the center of
           the domain. Black lines are magnetic field lines.}
  \label{fig1}
\end{figure}

\begin{equation}
i(\nu)=C_1 K B_{\bot}^{\,\alpha+1}\nu^{\,-\alpha},
\label{radio}
\end{equation}

\noindent
where $C_1$ is a constant, $B_{\bot}$ is the component of the magnetic
field perpendicular to the line-of-sight (LoS), $\nu$ is the frequency
of the radiation, $\alpha=(\zeta-1)/2$ is the synchrotron spectral index
(assumed to be uniform everywhere and taken as 0.5 as observed in many
BSNRs). To compute the total radio intensity (Stokes parameter $I$)
at a given frequency $\nu_0$, we integrate the emissivity $i(\nu_0)$
along the LoS:

\begin{equation}
I(\nu_0) = \int i(\nu_0) ~dl~,
\end{equation}

\noindent
where $dl$ is the increment along the LoS.

The normalization of the electron distribution $K\rs{s}$ in
Eq. \ref{radio} (index ``s'' refers to the immediately post-shock values)
depends on the injection efficiency (the fraction of electrons that
move into the cosmic-ray pool). Unfortunately, it is unknown how the
injection efficiency evolves in time.  On theoretical grounds, $K_{\rm s}$
is expected to vary with the shock velocity $V\rs{sh}(t)$ and, in case
of inhomogeneous ISM, with the immediately post-shock value of mass
density, $\rho\rs{s}$; let us assume that approximately $K\rs{s}\propto
\rho\rs{s}V\rs{sh}(t)^{-b}$. \cite{1998ApJ...493..375R} considered
three empirical alternatives for $b$ as a free parameter, namely,
$b=0,-1,-2$. \cite{pet-band-2006} showed that one can expect $b>0$ and
its value could be $b\approx 4$. Stronger shocks are more successful
in accelerating particles. To be accelerated effectively, a particle
should obtain in each Fermi cycle larger increase in momentum, which is
proportional to the shock velocity. Negative $b$ reflects an
expectation that {\it injection} efficiency may behave in a way similar
to {\it acceleration} efficiency: stronger shocks might inject particles
more effectively. In contrast, positive $b$ represents a different point
of view: efficiencies of injection and acceleration may have opposite
dependencies on the shock velocity. Stronger shock produces higher
turbulence which is expected to prevent more thermal particles to recross
the shock from downstream to upstream and to be, therefore, injected.
Since the picture of injection is quite unclear from both theoretical
and observational points of view, we do not pay attention to the
physical motivations of the value of $b$. Instead, our goal is to see
how different trends in evolution of injection efficiency may affect
the visible morphology of SNRs. Such understanding could be useful for
future observational tests on the value of $b$.

We found, in agreement with \cite{1998ApJ...493..375R}, that the value
of $b$ does not affect the main features of the surface brightness
distribution if SNR evolves in uniform ISM.  Therefore we use the
value $b=0$ to produce the SNR images in models with uniform ISM
(models DZ1, DZ2, DX1, and DX2). In cases where non-uniformity of ISM
causes variation of the shock velocity in SNR (models GZ1, GZ2, GX1,
and GX2), we calculate images for $b=-2,0,2$. We follow the model of
\cite{1998ApJ...493..375R} in description of the post-shock evolution
of relativistic electrons. Adopting this approach and considering that
$\zeta=2$ (being $\alpha = 0.5$, see above), one obtains that (see
Appendix \ref{app1})

\begin{equation}
 \frac{K(a,t)}{K\rs{s}(R,t)}=
 \left(\frac{P(a,t)}{P(R,t)}\right)^{-b/2}
 \left(\frac{\rho\rs{o}(a)}{\rho\rs{o}(R)}\right)^{-(b+1)/3}
 \left(\frac{\rho(a,t)}{\rho(R,t)}\right)^{5b/6+4/3}
 \label{K-evol}
\end{equation}

\noindent
where $a$ is the lagrangian coordinate, $R$ is the shock radius, $\rho$
is the gas density, $P$ is the gas pressure, and the index ``o'' refers to
the pre-shock values. It is important to note that this formula accounts
for variation of injection efficiency caused by the non-uniformity of ISM.

The electron injection efficiency may also vary with the obliquity
angle between the external magnetic field and the shock normal,
$\phi\rs{Bn}$. The numerical simulations suggest that injection
efficiency is larger for parallel shocks, i.e. where the magnetic
field is parallel to the shock speed (obliquity angle close to
zero; \citealt{1995ApJ...453..873E}). However, it has been shown
(\citealt{1990ApJ...357..591F}) that models with injection strongly
favoring parallel shocks produce SNR maps that do not resemble
any known objects (it is also claimed that injection is more
efficient where the magnetic field is perpendicular to the shock speed;
\citealt{1987ApJ...313..842J}). On the other hand, comparison of known
SNRs morphologies with model SNR images calculated for different strengths
of the injection efficiency dependence on obliquity suggests that the
injection efficiency in real SNRs could not depend on obliquity (Petruk,
in preparation). In such an unclear situation, we consider the three
cases: quasi-parallel, quasi-perpendicular, and isotropic injection
models. Following \cite{1990ApJ...357..591F}, we model quasi-parallel
injection by multiplying the normalization of the electron distribution
$K$ by $\cos^2 \phi\rs{Bn2}$ (see also \citealt{1989ApJ...338..963L}),
where $\phi\rs{Bn2}$ is the angle between the shock normal and the
post-shock magnetic field\footnote{For a shock compression ratio of
4 (the shock Mach number is $\gg 10$ in all directions during the whole
evolution in each of our simulations), the obliquity angle between the
external magnetic field and the shock normal, $\phi\rs{Bn}$, is related to
$\phi\rs{Bn2}$ by $\sin^2 \phi\rs{Bn2} = (\cot^2 \phi\rs{Bn}/16+1)^{-1}$
(e.g. \citealt{1990ApJ...357..591F}).}. By analogy with the quasi-parallel
case, we model quasi-perpendicular injection by multiplying $K$ by
$\sin^2 \phi\rs{Bn2}$.

An important point is the degree of ordering of magnetic field
downstream of the shock. Radio polarization observation of a
number of SNRs (e.g. Tycho \citealt{1991AJ....101.2151D}, SN1006
\citealt{1993AJ....106..272R}) show the low degree of polarization,
10-15\% (in case of ordered magnetic field the value expected is about
70\%; \citealt{1990ApJ...357..591F}), indicating highly disordered
magnetic field. Thus we calculate the synchrotron images of SNR for
two opposite cases. First, since our MHD code gives us the three
components of magnetic field, we are able to calculate images with
ordered magnetic field. Second, we introduce the procedure of
the magnetic field disordering (with randomly oriented magnetic field
vector with the same magnitude in each point) and then synthesize
the radio maps. In models which have a disordered magnetic field, we
use the post-shock magnetic field before disordering to calculate the
angle $\phi\rs{Bn2}$; as discussed by \cite{1990ApJ...357..591F}, this
corresponds to assume that the disordering process takes place over a
longer time-scale than the electron injection, occurring in the close
proximity of the shock. Since we found that the asymmetries induced by
gradients either of ambient plasma density or of ambient magnetic field
strength are not significantly affected by the degree of ordering of the
magnetic field downstream of the shock, in the following we will focus
on the models with disordered magnetic field.

The goal of this paper is to look whether non-uniform ISM or non-uniform
magnetic field can produce asymmetries on BSNRs morphology. In order to
clearly see the role of these two factors in determining the morphology
of BSNRs, we use some simplifying assumptions about electron kinetic
and behavior of magnetic field in vicinity of the shock front. Our
calculations are performed in the test-particle limit, i.e. they ignores
the energy in cosmic rays. In particular, we do not consider possible
amplification of magnetic field by the cosmic-ray streaming instability
(\citealt{2000MNRAS.314...65L}, \citealt{2001MNRAS.321..433B}). We
expect that the main features of the modeled SNR morphology will not
change if this process is independent of obliquity angle. If future
investigations show undoubtedly that magnetic field amplification varies
strongly with obliquity, the role of this effect in producing BSNRs have
to be additionally studied.

\section{Results}
\label{sec3}

\begin{figure}[!t]
  \centering
  \includegraphics[width=8.5cm]{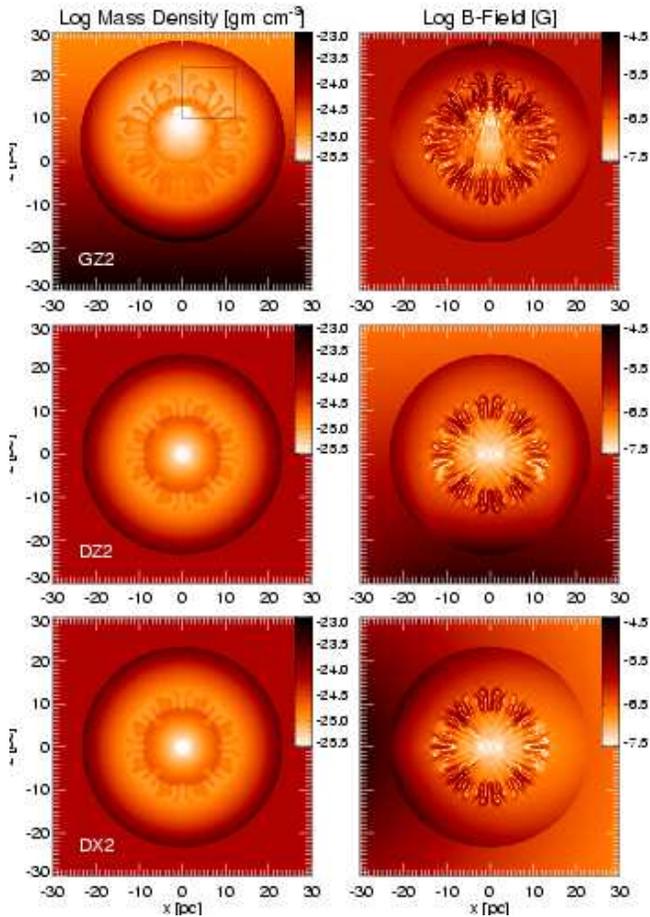}
  \caption{2-D sections in the $(x,z)$ plane of the mass density
   distribution (left panels), in log scale, and of the distribution
   of the magnetic-field strength (right panels), in log scale, in the
   simulations GZ2 (upper panels), DZ2 (middle panels), and DX2 (lower
   panels) at $t=18$ kyrs. The box in the upper left panel marks the
   region shown in Fig.~\ref{fig3}.}
\label{fig2}
\end{figure}

\begin{figure}[!t]
  \centering
  \includegraphics[width=8.5cm]{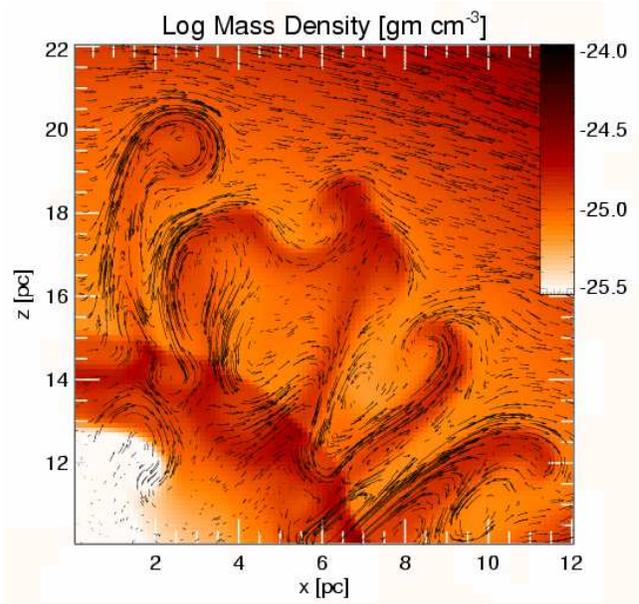}
  \caption{Close-up view of the region marked with a box in
           Fig.~\ref{fig2}. The dark fingers mark the R-M instability.
           The magnetic field is described by the superimposed arrows
           the length of which is proportional to the magnitude of the
           field vector.}
\label{fig3}
\end{figure}

In all the models examined, we found the typical evolution of
adiabatic SNRs expanding through an organized ambient magnetic field
(see \citealt{2001ApJ...563..800B} and references therein): the fast
expansion of the shock front with temperatures of few millions degrees,
and the development of Richtmyer-Meshkov (R-M) instability, as the forward
and reverse shocks progress through the ISM and ejecta, respectively
(see \citealt{1999ApJ...511..335K}). As examples, Fig.~\ref{fig2} shows
2-D sections in the $(x,z)$ plane of the distributions of mass density
and of magnetic field strength for the models GZ2, DZ2, and DX2 at $t=18$
kyrs. The inner shell is dominated by the R-M instability that causes the
plasma mixing and the magnetic field amplification. In the inner shell,
the magnetic field shows a turbulent structure with preferentially
radial components around the R-M fingers (see Fig.~\ref{fig3}). Note
that, some authors have invoked the R-M instabilities to explain the
dominant radial magnetic field observed in the inner shell of SNRs
(e.g. \citealt{1996ApJ...472..245J}); however, in our simulations, the
radial tendency is observed well inside the remnant and not immediately
behind the shock as inferred from observations.

We found that, throughout the expansion, the shape of the remnant
is not appreciably distorted by the ambient magnetic field because,
for the values of explosion energy and ambient field strength (typical
of SNRs) used in our simulations, the kinetic energy of the shock is
many orders of magnitude larger than the energy density in the ambient
\mag field (see also \citealt{1990ApJ...355L..47M}). The shape of the
remnant does not differ visually from a sphere also in the cases with
density stratification of the ambient medium\footnote{In these cases,
the remnant appears shifted toward the low density region; see upper
panels in Fig.~\ref{fig2} (see also \citealt{1996ApJ...471..279D}).}
as it is shown by \cite{1999A&A...344..295H}.

The radio emission of the evolved remnants is characterized by
an incomplete shell morphology when the viewing angle is not
aligned with the direction of the average ambient magnetic field
(cf. \citealt{1990ApJ...357..591F}); in general, the radio emission shows
an axis of symmetry with low levels of emission along it, and two bright
limbs (arcs) on either side (see also \citealt{1998ApJ...493..781G}). This
morphology is very similar to that observed in BSNRs.

\subsection{Obliquity angle dependence}
\label{obl-ang}

\begin{figure*}[!ht]
  \centering
  \includegraphics[width=16.cm]{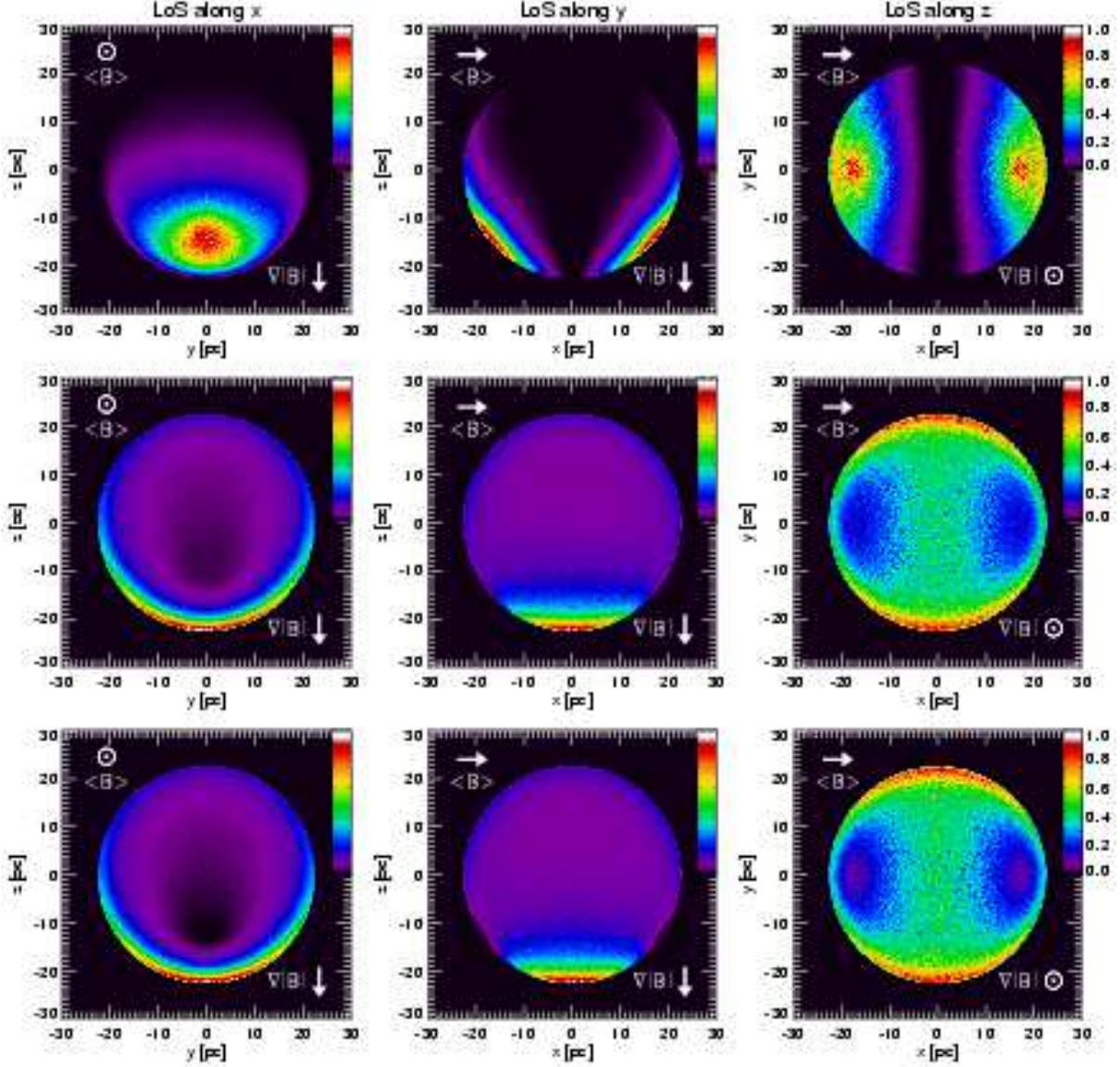}
  \caption{Synchrotron radio emission (normalized to the maximum of
           each panel), at $t=18$ kyrs, synthesized from model
           DZ1 assuming $b=0$ (see text) and randomized internal
           magnetic field, when the LoS is aligned with the $x$
           (left), $y$ (center), or $z$ (right) axis. The figure
           shows the quasi-parallel (top), isotropic (middle), and
           quasi-perpendicular (bottom) particle injection cases. The
           color scale is linear and is given by the bar on the right.
           The directions of the average unperturbed ambient magnetic
           field, {\avemag}, and of the magnetic field strength gradient,
           $\nabla |\mag|$, are shown in the upper left and lower right
           corners of each panel, respectively.}
\label{fig4} \end{figure*}

For each of the models listed in Table \ref{tab1}, we synthesized
the synchrotron radio emission, considering each of the three cases
of variation of electron injection efficiency with shock obliquity:
quasi-parallel, quasi-perpendicular, and isotropic particle injection.
As an example, Fig.~\ref{fig4} shows the synchrotron radio emission
synthesized from the uniform ISM model DZ1 with randomized internal
magnetic field at $t=18$ kyrs in each of the three cases. We recall that
for these uniform density cases, we have adopted an injection efficiency
independent from the shock speed ($b=0$, Sect. \ref{e-acc}). All
images are maps of total intensity normalized to the maximum intensity
of each map and have a resolution of 400 beams per remnant diameter
($D\rs{SNR}$). The images are derived when the LoS is parallel to the
average direction of the unperturbed ambient magnetic field {\avemag}
(LoS aligned with the $x$ axis), or perpendicular both to {\avemag}
and to the gradient of field strength (LoS along $y$), or parallel to
the gradient of field strength (LoS along $z$).

The different particle injection models produce images that can differ
considerably in appearance. In particular, the quasi-parallel case leads
to morphologies of the remnant not reproduced by the other two cases: a
center-brightened SNR when the LoS is aligned with $x$ (top left panel
in Fig.~\ref{fig4}), a BSNR with two bright arcs slanted and converging on
the side where \mag field strength is higher when the LoS is along $y$
(top center panel), and a remnant with two symmetric bright spots
located between the center and the border of the remnant when the LoS
is along $z$ (top right panel). Neither the center-brightened remnant
nor the double peak structure, showing no structure describable as
a shell, seems to be observed in SNRs\footnote{Excluding filled center
and composite SNRs, but these are due to energy input from a central
pulsar.}. We found analogous morphologies in all the models listed in
Table \ref{tab1}, considering the quasi-parallel case. As extensively
discussed by \cite{1990ApJ...357..591F} for models with uniform ambient
magnetic field and $b=-2$, we also conclude that the quasi-parallel
case leads to radio images unlike any observed SNR (see also
\citealt{1987A&A...183..118K}).

The isotropic case leads to remnant's morphologies similar to those
produced in the quasi-perpendicular case although the latter case shows
deeper minima in the radio emission than the first one. When the LoS is
aligned with $x$ (middle left and bottom left panels in Fig.~\ref{fig4})
or with $y$ (middle center and bottom center panels), the remnants have
one bright arc on the side where the \mag strength is higher. When the
LoS is aligned with $z$ (middle right and bottom right panels), the
remnants have two opposed arcs that appear perfectly symmetric. We found
that the isotropic and quasi-perpendicular cases lead to morphologies
of the remnants similar to those observed.

\subsection{Non-uniform ISM: dependence from parameter $b$}

\begin{figure*}[!ht]
  \centering
  \includegraphics[width=16.cm]{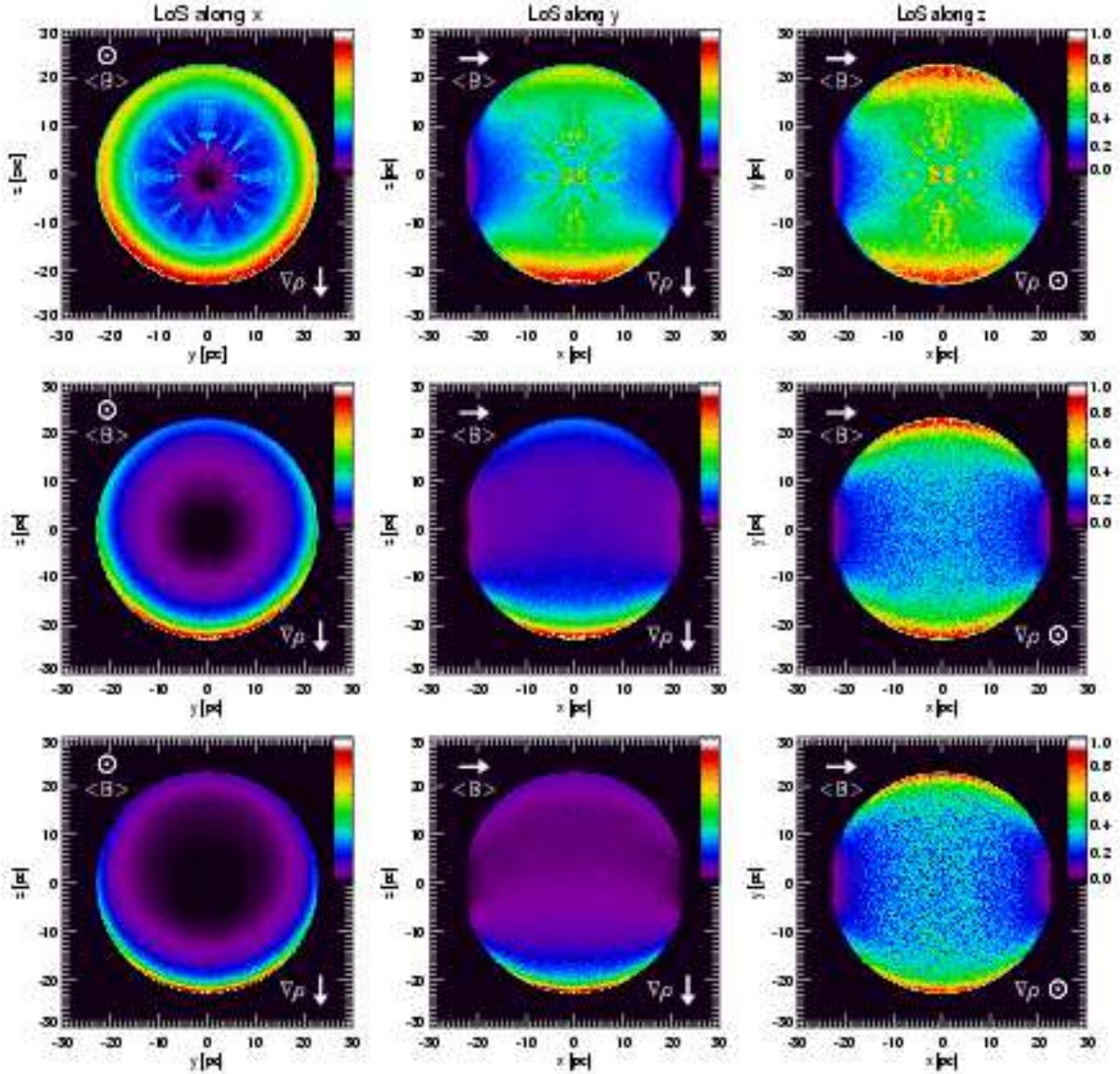}
  \caption{Presentation as in Fig.~\ref{fig4} for model GZ1 with
           randomized internal magnetic field, assuming
           quasi-perpendicular particle injection and $b=-2$ (top panels),
           $b=0$ (middle) and $b=2$ (bottom). The directions of the
           average unperturbed ambient magnetic field, {\avemag},
           and of the ambient plasma density gradient, $\nabla \rho$,
           are shown in the upper left and lower right corners of each
           panel, respectively.}
\label{fig5}
\end{figure*}

For models describing the SNR expansion through a non-uniform ISM (models
GZ1, GZ2, GX1, GX2), we derived the synthetic radio maps considering
three alternatives for the dependence of the injection efficiency on the
shock speed, namely $b=-2,0,2$ (see Sect. \ref{e-acc}). As an example,
Fig.~\ref{fig5} shows the synthetic maps derived from model GZ1 with
randomized internal magnetic field, assuming quasi-perpendicular particle
injection, and considering $b=-2$ (top panels), $b=0$ (middle) and
$b=2$ (bottom).

When the LoS is not aligned with the density gradient, the radio images
show asymmetric morphologies of the remnants. In this case, the main
effect of varying $b$ is to change the degree of asymmetry observed in
the radio maps. In the example shown in Fig.~\ref{fig5}, the density
gradient is aligned with the $z$ axis and asymmetric morphologies are
produced when the LoS is aligned with $x$ (left panels) or with $y$
(center panels). In all the cases, the remnant is brighter where the
mass density is higher. On the other hand, the degree of asymmetry
increases with increasing value of $b$.

The reason of such behavior consists in the balance between the roles
of the shock velocity and of density in changing the injection efficiency.
Consider, as an example, the top left panel in Fig.~\ref{fig5}: the
increase of the shock velocity on the north (due to fall of the ambient
density) leads to an increase of the brightness there (due to rise of
the injection efficiency) that partially balances the increase of the
brightness on the south due to higher density of ISM. On the other hand,
for the model shown in the bottom left panel in Fig.~\ref{fig5}, the
fraction of accelerated electrons increases on the south due to both
the rise of density and the decrease of the shock velocity.
\begin{figure*}[!ht]
  \centering
  \includegraphics[width=16.cm]{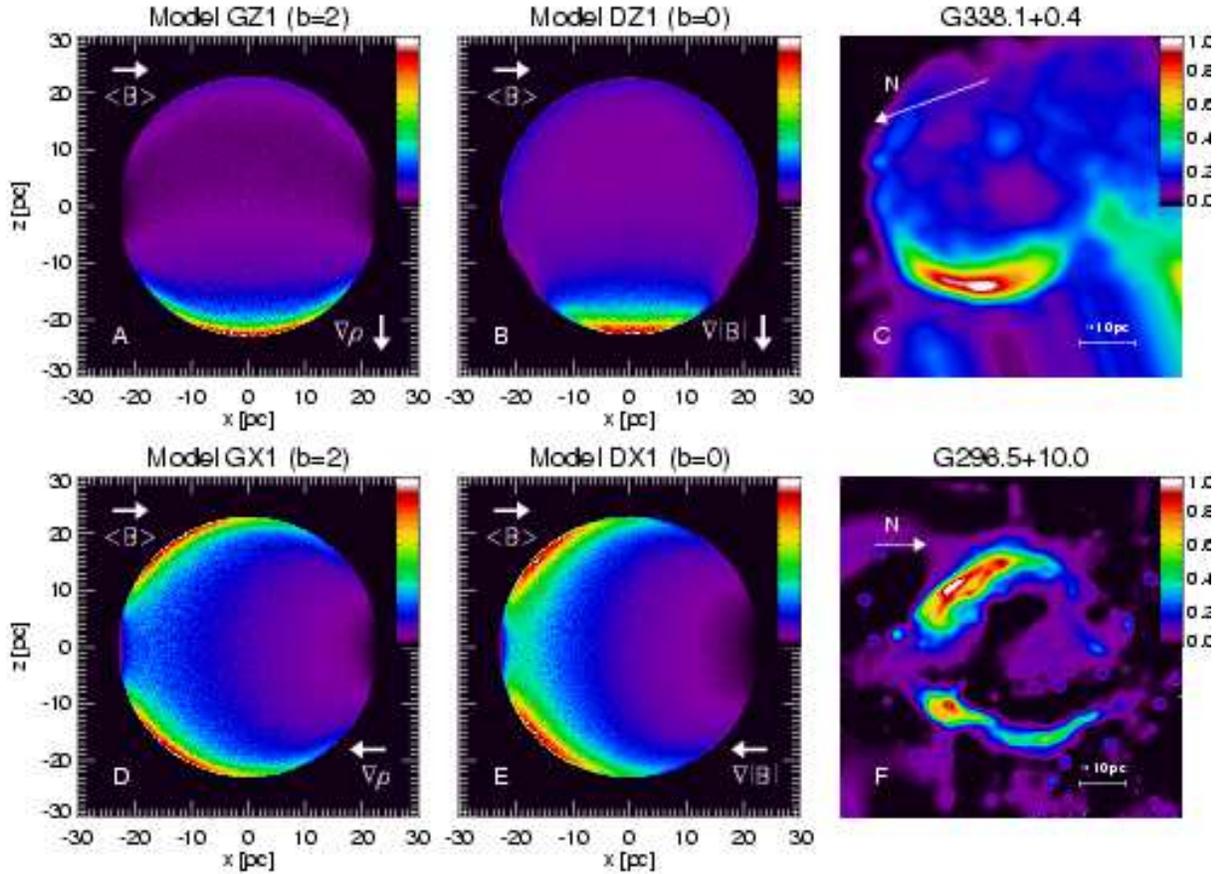}
  \caption{Synchrotron radio emission (normalized to the maximum of
           each panel), at $t=18$ kyrs, synthesized from models
           assuming a gradient of ambient plasma density (panels A and
           D; with $b=2$) or of ambient magnetic field strength (panels
           B and E; with $b=0$) when the LoS is aligned with the $y$
           axis. All the models assume quasi-perpendicular particle
           injection. The directions of the average unperturbed
           ambient magnetic field, {\avemag}, and of the plasma
           density or magnetic field strength gradient, are shown
           in the upper left and lower right corners of each panel,
           respectively. The right panels show two examples of
           radio maps (data adapted from \citealt{1996A&AS..118..329W}
           and \citealt{1998ApJ...493..781G}; the arrows point in the
           north direction) collected for the SNRs G338.1+0.4 (panel C)
           and G296.5+10.0 (panel F). The color scale is linear and is
           given by the bar on the right.}
\label{fig6}
\end{figure*}

When the LoS is aligned with the density gradient, the radio images are
symmetric. In the example shown in Fig.~\ref{fig5}, this corresponds to
the maps derived when the LoS is along $z$ (right panels); the remnants
are characterized by two opposed arcs with identical surface brightness.

\subsection{Morphology}

Fig.~\ref{fig6} shows the radio emission maps, at a time of 18 kyrs,
synthesized from models with a gradient of ambient plasma density (panels
A and D; assuming $b=2$) and of ambient \mag field strength (panels
B and E; assuming $b=0$). All the models assume quasi-perpendicular
particle injection (the isotropic case produces radio maps with similar
morphologies and the quasi-parallel case is discussed later) and
randomized internal magnetic field. The viewing angle is perpendicular
both to the average direction of the unperturbed ambient magnetic field,
{\avemag}, (direct along the $x$ axis) and to the gradients of density
or field strength (direct either along $z$, panels A and B, or $x$,
panels D and E). The right panels show, as examples, the radio maps of
the SNRs G338.1+04 (panel C, data from \citealt{1996A&AS..118..329W})
and G296.5+10.0 (panel F, from \citealt{1998ApJ...493..781G}).

In the quasi-perpendicular case discussed here, the maximum synchrotron
emissivity is reached where the magnetic field is strongly compressed.
This configuration has been referred as ``equatorial belt''
(e.g. \citealt{2004A&A...425..121R}); {\avemag} runs between the two
opposed arcs (along the $x$ axis). We found that, when the density or the
magnetic field strength gradient is perpendicular to the field itself,
the morphology of the radio map strongly depends on the viewing angle. In
these cases, the two opposed arcs appear perfectly symmetric when the LoS
is aligned with the gradient (see, for instance, the right panels in
Fig.~\ref{fig5}), otherwise the two arcs can have very different radio
brightness, leading to strongly asymmetric BSNRs (see panels A and B
in Fig.~\ref{fig6}). In the former case (LoS aligned with the gradient),
the remnant is characterized by two axes of symmetry: one between the
two symmetric arcs and the other perpendicular to the two. In models
with strong magnetic field strength gradients (DZ2; \mag varies by
a factor $\sim 60$ over 60 pc), we found that the radio images are
center-brightened when the LoS is aligned with the gradient (figure not
reported). The fact that center-brightened remnants are not observed
suggests that the external \mag varies moderately in the neighborhood
of the remnants.

In case of asymmetry, the gradient is always perpendicular to the arcs,
and the brightest arc is located where either magnetic field strength
or plasma density is higher (see panels A and B in Fig.~\ref{fig6}),
since the synchrotron emission depends on the plasma density, on
the pressure, and on the field strength (see Eqs. \ref{radio} and
\ref{K-evol}); in this case, there is only one axis of symmetry oriented
along the density or \mag gradient. When the LoS is parallel to {\avemag}
(along $x$ in our models), the radio maps show a shell structure with a
maximum intensity located where magnetic field strength or plasma density
is higher (see left panels in Fig.~\ref{fig4} for isotropic and
quasi-perpendicular cases and left panels in Fig.~\ref{fig5}). Our
simulations show that, when the density or the magnetic field strength
gradient is perpendicular to the field itself, remnants with a monopolar
morphology can be observed at LoS not aligned with the gradient (see
also \citealt{1990ICRC....4...72R}). Examples of observed monopolar
remnants are G338.1+0.4 (see panel C in Fig.~\ref{fig6}) or G327.4+1.0
or G341.9-0.3.
\begin{figure*}[!ht]
  \sidecaption
  \includegraphics[width=12.cm]{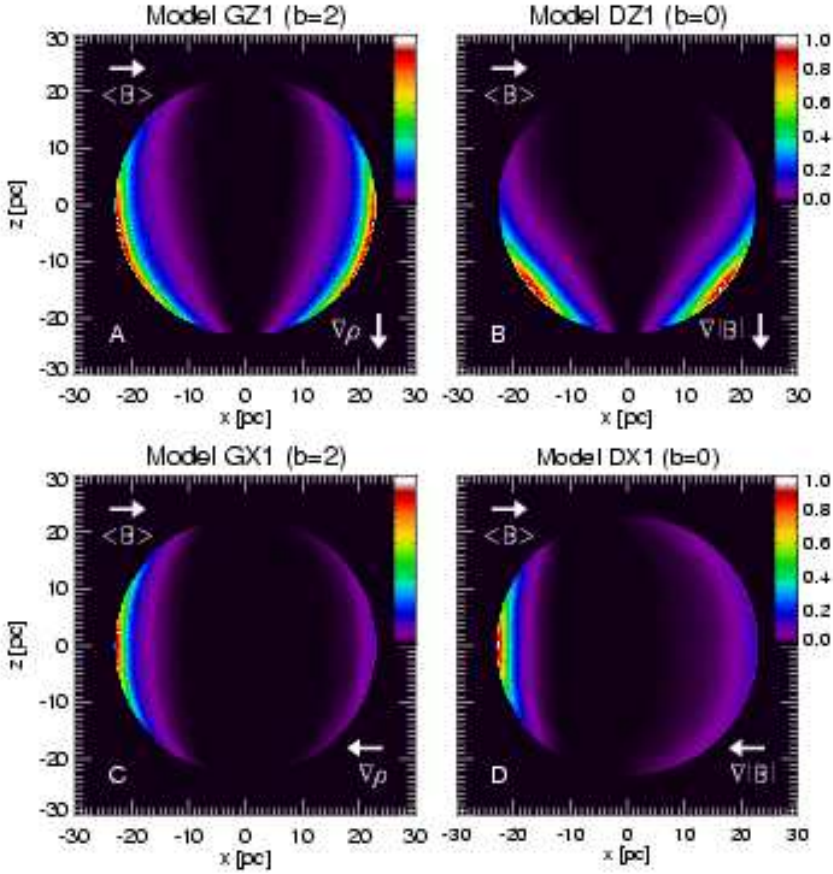}
  \caption{Presentation as in Fig.~\ref{fig6}, assuming quasi-parallel
           instead of quasi-perpendicular particle injection.}
\label{fig7}
\end{figure*}

When the density or \mag field strength gradient is parallel to {\avemag}
(panels D and E in Fig.~\ref{fig6}) and the LoS lies in the plane
perpendicular to {\avemag}, the morphology of the radio map does not
depend on the viewing angle and the two opposed arcs have the same radio
brightness. In these cases, however, there is only one axis of symmetry
and the two arcs appear slanted and converging on the side where field
strength or plasma density is higher; again, the symmetry axis is aligned
with the density or \mag strength gradient. Examples of this kind of
objects are G296.5+10.0 (see panel F in Fig.~\ref{fig6}) or G332.4-004 or
SN1006 (which is, however, much younger than the simulated SNRs). When the
external magnetic field is parallel to the LoS, because the system is
symmetric about the magnetic field, the remnant is axially symmetric
and the radio maps show a complete radio shell at constant intensity.

In the quasi-parallel case, {\avemag} runs across the arcs. This
configuration has been referred as ``polar caps'' and it has been
invoked for the SN1006 remnant (\citealt{2004A&A...425..121R}). The
quasi-parallel case, apart from the center-brightened morphology discussed
in Sect. \ref{obl-ang}, can also produce remnant morphologies similar to
those shown in Fig.~\ref{fig6}. As examples, Fig.~\ref{fig7} shows the
radio emission maps obtained in the cases discussed in Fig.~\ref{fig6}
but assuming quasi-parallel instead of quasi-perpendicular particle
injection. Again, the viewing angle is perpendicular both to {\avemag}
(direct along the $x$ axis) and to the gradients of density or field
strength (direct either along $z$, panels A and B, or $x$, panels C and
D). In the quasi-parallel case, remnants with a bright radio limb are
produced if the gradient of ambient density or of ambient \mag field
strength is parallel to {\avemag} (instead of perpendicular to {\avemag}
as in the quasi-perpendicular case), whereas slanting similar radio arcs
are obtained if the gradient is perpendicular to {\avemag} (instead of
parallel as in the quasi-perpendicular case).

\section{Discussion}
\label{sec4}

Our simulations show that asymmetric BSNRs are explained if the ambient
medium is characterized by gradients either of density or of ambient
magnetic field strength: the two opposed arcs have different surface
brightness if the gradient runs across the arcs (see panels A and B in
Fig.~\ref{fig6}, and panels C and D in Fig.~\ref{fig7}), whereas the
two arcs appear slanted and converging on one side if the gradient runs
between them (see panels D and E in Fig.~\ref{fig6} and panels A and B in
Fig.~\ref{fig7}). In all the cases (including the three alternatives for
the particle injection), the symmetry axis of the remnant is always
aligned with the gradient.

From the radio maps, we derived the azimuthal intensity profiles: we
first find the point on the map where the intensity is maximum; then the
contour of points at the same distance from the center of the remnant
as the point of maximum intensity defines the azimuthal radio intensity
profile. Following \cite{1990ApJ...357..591F}, we quantify the degree of
``bipolarity'' of the remnants by using the so-called azimuthal intensity
ratio $A$, i.e. the ratio of maximum to minimum intensity derived from
the azimuthal intensity profiles. In addition, we quantify the degree of
asymmetry of the BSNRs by using a measure we call the azimuthal intensity
ratio $R\rs{max}\geq 1$, i.e. the ratio of the maxima of intensity
of the two limbs as derived from the azimuthal intensity profiles,
and the azimuthal distance $\theta\rs{D}$, i.e. the distance in deg
of the two maxima. In the case of symmetric BSNRs, $R\rs{max}=1$ and
$\theta\rs{D}=180^o$. As already noted by \cite{1990ApJ...357..591F},
the parameter $A$ depends on the spatial resolution of the radio maps and
on the aspect angle (i.e. the angle between the LoS and the unperturbed
magnetic field); moreover we note that, in real observations, the
measure of $A$ gives a lower limit to its real value if the background
is not accurately taken into account.  On the other hand, the parameters
$R\rs{max}$ and $\theta\rs{D}$ have a much less critical dependency on
these factors and, therefore, they may provide a more robust diagnostic
in the comparison between models and observations.
\begin{figure*}[!ht]
  \sidecaption
  \includegraphics[width=12cm]{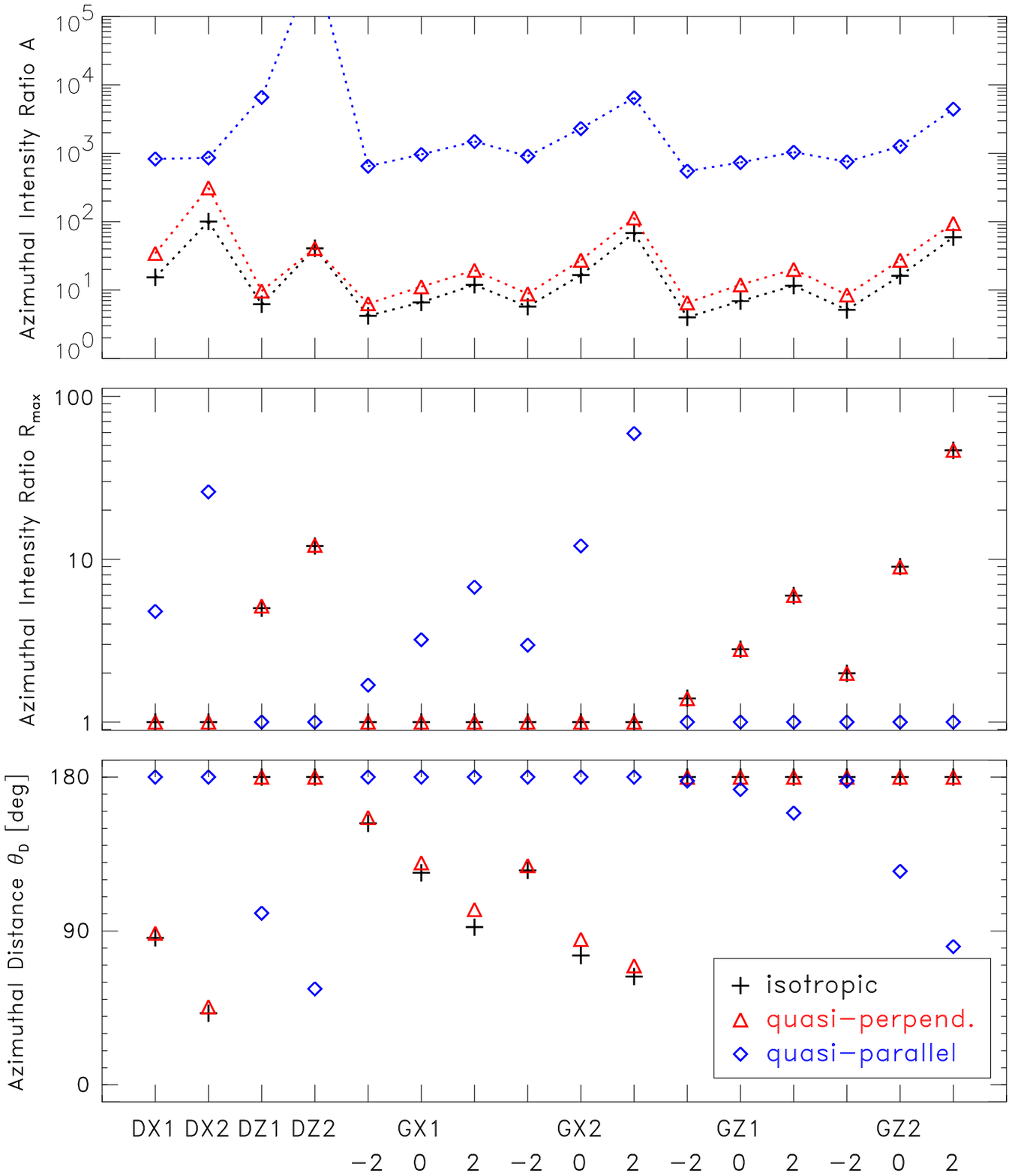}
  \caption{Azimuthal intensity ratio $A$ (i.e. the ratio of maximum to
           minimum intensity around the shell of emission - see text;
           upper panel), azimuthal intensity ratio $R\rs{max}$ (i.e. the
           ratio of the maxima of intensity of the two limbs around the
           shell; middle panel), and azimuthal distance $\theta\rs{D}$
           (i.e. the distance in deg of the two maxima of intensity around
           the shell; lower panel) for all the cases examined, considering
           the LoS aligned with the $y$ axis and a spatial resolution
           of 25 beams per remnant diameter, $D\rs{SNR}$. Black crosses:
           isotropic; red triangles: quasi-perpendicular; blue diamonds:
           quasi-parallel.}
\label{fig8} \end{figure*}

Fig.~\ref{fig8} shows the values of $A$, $R\rs{max}$, and $\theta\rs{D}$
derived for all the cases examined in this paper, considering the LoS
aligned with the $y$ axis, and radio maps with a resolution of 25 beams
per remnant diameter\footnote{After the radio maps are calculated, they
are convolved with a gaussian function with $\sigma$ corresponding to
the required resolution.} ($D\rs{SNR}$). Note that, our choice of the LoS
aligned with $y$ (aspect angle $\phi=90^o$) implies that the values of $A$
in Fig.~\ref{fig8} are upper limits, being $A$ maximum at $\phi=90^o$
and minimum at $\phi=0^o$ (see \citealt{1990ApJ...357..591F}). The
three models of particle injection (isotropic, quasi-perpendicular and
quasi-parallel) lead to different values of $A$. In the isotropic and
quasi-perpendicular cases, most of the values of $A$ range between 5 and
20 (for model DX2, $A$ is even larger than 100); in the quasi-parallel
case, the values of $A$ are larger than 500.

We found that, in general, a gradient of the ambient magnetic field
strength leads to remnant morphologies similar to those induced by a
gradient of plasma density (compare, for instance, panel A with B and
panel D with E in Fig.~\ref{fig6}). On the other hand, if $b<0$ in GX and
GZ models, ambient \mag field gradients are more effective in determining
the morphology of asymmetric BSNRs. This is seen in a more quantitative
form in Fig.~\ref{fig8}. DX and DZ models give $R\rs{max}$ values
higher and $\theta\rs{D}$ values lower than GX and GZ models with $b <
0$: a modest gradient of the magnetic field (models DX1 and DZ1) gives
a value of $R\rs{max}$ higher or $\theta\rs{D}$ lower than the two models
with strong density gradients (models GX2 and GZ2) and $b < 0$.

Fig.~\ref{fig8} also shows that, in models with a density gradient,
the degree of asymmetry of the remnant increases with increasing value
of $b$; the GX and GZ models with $b>0$ give values of $R\rs{max}$
and $\theta\rs{D}$ comparable with (or, in the case of $R\rs{max}$,
even larger of) those derived from DX and DZ models. In the case of
quasi-parallel particle injection for remnants with converging similar
arcs, it is necessary a strong gradient of density perpendicular to
\mag and $b\ge 0$ (compare models GZ1 and GZ2 in the lower panel in
Fig.~\ref{fig8}) to give values of $\theta\rs{D}$ comparable to those
obtained with a moderate gradient of ambient \mag field strength
perpendicular to \mag (see model DZ1 in Fig.~\ref{fig8}).

In order to compare our model predictions with observations of real
BSNRs, we have selected 11 SNR shells which show one or two clear lobes
of emission in archive total intensity radio images, separated by a
region of minima. We have discarded all those cases in which several
point-like or extended sources appear superimposed to the bright
limbs, or other cases in which the location of maximum or minimum
emission around the shell is difficult to derive. Unlike other lists of
BSNRs published in the literature (e.g. \citealt{1987A&A...183..118K,
1990ApJ...357..591F, 1998ApJ...493..781G}), here we focus on a reliable
measure of the parameters $A$, $R\rs{max}$ and $\theta\rs{D}$;
we avoid, therefore, patchy and irregular limbs, as in the case of
G320.4-01.2 of \citet{1998ApJ...493..781G}. Moreover, we are obviously
not discarding remnants which have constraints on $A$, $R\rs{max}$
or $\theta\rs{D}$ (e.g. \citealt{1990ApJ...357..591F} considered only
cases with $R\rs{max}<2$), and we are considering remnants observed with
a resolution greater than 10 beams per remnant diameter. Since in
our models we follow the remnant evolution during the adiabatic phase,
we also need to discard objects that are clearly in the radiative
phase. Unfortunately, for most of the objects selected, there is no
indication of their evolutionary stage in literature. Assuming that the
remnant expands in a medium with particle number density $n\rs{ism}\lsim
0.3$ cm$^{-3}$, the shock radius derived from the Sedov solution at
time $t\rs{tr}$ (i.e. at the transition time from the adiabatic to the
radiative phase; see Eq. \ref{trans_time}) is

\begin{equation}
r\rs{tr} = 19\;E\rs{51}^{5/17}\;n\rs{ism}^{-7/17}\lsim 35~\mbox{pc}~,
\end{equation}

\noindent
where we have assumed that $E\rs{51} = 1.5$. Therefore, we only considered
remnants with radius $r\rs{snr} < 35$ pc (i.e. with size $< 70$~pc) that
are, most likely, in the adiabatic phase. Our list does not pretend to be
complete or representative of the class, and it is compiled to derive the
observed values of the parameters $A$, $R\rs{max}$ and $\theta\rs{D}$
with the lowest uncertainties. For this reason, we have considered
remnants for which a total intensity radio image in digital format is
available. Actually, in most of the cases, we have used the 843 MHz data
of the MOST supernova remnant catalogue (\citealt{1996A&AS..118..329W}).
\begin{table*}[!ht]
\caption{List of barrel-shaped SNR shells for which a measurement of
$A$, $R\rs{max}$, and $\theta\rs{D}$ is presented for comparison with
our models.}
\label{tab2}
\medskip \begin{minipage}{18cm}
\begin{center}
\begin{tabular}{lcccccccccl}
\hline
\hline
Remnant\footnote{All the data are from the MOST supernova remnant catalogue
(\citealt{1996A&AS..118..329W}), except where noted.} & Flux & size   &
d   & size & Res.\footnote{Spatial resolution of the observation in
beams per remnant diameter.} &$A$  &  $R\rs{max}$  & $\theta\rs{D}$ & Ref./notes\\
    & Jy   & arcmin & kpc & pc   & beams/$D\rs{SNR}$ &    &     &  deg  & \\
\hline
\multicolumn{10}{c}{Evolved Remnants} \\
\hline
G296.5+10.0&48 &$90\times65$&$2.1$ &$55\times40$ &108 &$>11$ &$1.2$&$85$& 1 \\
G299.6-0.5 &1.1&$13\times13$&$18.1$&$68$         &18 &$6$   &$2$ &$160$& 2  \\
G304.6+0.1 &18 & $8\times8$ &$7.9$ &$18$         &11 &$20$ &$1.5$&$120$& 3 \\
G327.4+1.0 &2.1&$14\times13$&$13.9$&$56$         &19 &$>10$ &$>10$&ND & 2,4 \\
G332.0+0.2 &8.9&$12\times12$&$<20$ &$<70$        &17 &$5$   &$1$ &$145$& 2,7\\
G338.1+0.4 &3.8&$16\times14$&$9.9$ &$46\times40$ &21 &$3$ &$2$ &$>120$& 2 \\
G341.9-0.3 &2.7& $7\times7$ &$14.0$&$28$         &10 &$8$   &$3$ &$170$& 2 \\
G346.6-0.2 &8.7&$11\times10$&$8.2$ &$26\times23$ &15 &$2$ &$1.1$&$110$& 2,7 \\
G351.7+0.8 &11 &$18\times14$&$6.7$ &$35\times27$ &22 &$2$ &$1.6$&$130$& 2  \\
\hline
\multicolumn{10}{c}{Young Remnants} \\
\hline
G327.6+14.6&19 &$30\times30$&$2.2$ &$19\times19$ &42 &$22$  &$1$ &$127$& 5 \\
G332.4-0.4 &34 &$11\times10$&$3.1$ &$10\times9$  &15 &$7$   &$1.6$&$98$ & 6 \\
\hline\hline
\smallskip
\end{tabular}
\end{center}
References and notes. - (1) A.k.a. PKS 1209-51/52. Age: 3--20 kyrs,
\citet{rmk88}. Distance from \citet{gdg00}. (2) Distance derived by
\citet{cb98} using a revised $\Sigma-D$ relation. (3) Distance from
\citet{cmr75}. (4) This shell has only one limb (``monopolar'' according
to the definition of \citealt{1990ApJ...357..591F}). $A$ and $R\rs{max}$
are lower limits and no $\theta\rs{D}$ is derived. (5)
A.k.a. SN1006. Distance from \cite{wgl03}. (6) A.k.a. RCW103. Distance
from \cite{rgj04}. (7) Two maxima have been found in one
lobe. $\theta\rs{D}$ is the average of the two.
\end{minipage}
\end{table*}

Our list is reported in Table \ref{tab2}. We have separated evolved and
young SNRs. While the young SNRs listed in Table \ref{tab2} have very
reliable measurement of $A$, $R\rs{max}$ and $\theta\rs{D}$ and a good record
of literature, making them very good candidate to test the diagnostic
power of our model, we stress that the models we are considering in this
paper are focused on evolved SNRs; we leave the discussion about young
SNRs to a separate work. For each object in Table \ref{tab2}, we show
the apparent size, the distance (from dedicated studies where possible,
otherwise from the revised $\Sigma-D$ relation of \citealt{cb98}; see
their paper for caveats on usage of the $\Sigma-D$ relation to derive
SNR distance), the real size, the resolution of the observation, and the
parameters $A$, $R\rs{max}$, and $\theta\rs{D}$ we have introduced here.

Table \ref{tab2} shows that most of the 11 remnants have $A\leq 10$,
i.e. values consistent with those derived in Fig.~\ref{fig8} for the three
alternatives for the particle injection (recall that the values shown
in the figure have to be considered as upper limits). Four remnants show
high values of $A$ ($10<A<100$) that are difficult to explain in terms of
the isotropic or the quasi-perpendicular injection models with $b < 0$
unless the remnant expands through a non-uniform ambient magnetic field
(see models DX2, and DZ2 in Fig.~\ref{fig8}). In the light of these
considerations, we cannot exclude a priori any of the three alternative
models for the particle injection.

Four of the 11 objects in Table \ref{tab2} show values of $R\rs{max}\geq
2$, pointing out that, in these objects, the bipolar morphology
is asymmetric with the two radio limbs differing significantly in
intensity. An example of this kind of remnants is G338.1+0.4 (see panel
C in Fig.~\ref{fig6}). In the light of our results, its morphology can be
explained if a gradient of ambient density or of ambient magnetic field
strength is either perpendicular to the average ambient magnetic field,
\avemag, in the isotropic and quasi-perpendicular cases or parallel to
{\avemag} in the quasi-parallel case. It is worth noting that reveling
such a gradient from the observations may be a powerful diagnostic to
discriminate among the alternative particle injection models, producing
real advances in the understanding of the nonthermal physics of strong
shock waves.

An extreme example of a monopolar remnant with a bright radio limb is
G327.4+1.0 whose value of $R\rs{max}$ is larger than 10. Fig.~\ref{fig8}
shows that high values of $R\rs{max}$ can be easily explained as due
to non-uniform ambient magnetic field strength or to non-uniform
ambient density if $b>0$. We suggest that the morphology of G327.4+1.0
may give some hints on the value of $b$ (and, therefore, on the dependence
of the injection efficiency on the shock velocity) if the observations
show that the asymmetry is due to a non-uniform ambient medium through
which the remnant expands.

In Table \ref{tab2}, six of the 11 remnants (including the two young
remnants SN1006 and RCW103) have values of $\theta\rs{D}<140^o$, pointing
out that, in these objects, the two bright radio limbs appear slanted and
converging on one side. An example of this class of objects is G296.5+10.0
(a.k.a PKS 1209-51/52) shown in panel F in Fig.~\ref{fig6}. In this case,
the value of $\theta\rs{D} \sim 85^o$ derived from the observations may
be easily explained as due to a gradient of magnetic field strength either
parallel to {\avemag} in the isotropic and quasi-perpendicular cases or
perpendicular to {\avemag} in the quasi-parallel case. Models
with a gradient of ambient density cannot explain the low values of
$\theta\rs{D}$ found for G296.5+10.0 unless the gradients are strong (the
density should change by a factor $60$ over 60 pc) and the dependence
of the injection efficiency on the shock velocity gives\footnote{Large
positive values of $b$ do not necessarily mean an increasing fraction of
shock energy going into relativistic particles as the shock slows down
because decelerating shock accelerates particles to smaller $E\rs{max}$,
namely the maximum energy at which the electrons are accelerated.}
$b\geq 2$.

\section{Conclusions}
\label{sec5}

Our findings have significant implications on the diagnostics and lead
to several useful conclusions:

1. The three different particle injection models (namely, quasi-parallel,
quasi-perpendicular and isotropic dependence of injection efficiency
from shock obliquity) can produce considerably different images (see
Fig.~\ref{fig4}). The isotropic and quasi-perpendicular cases lead to
radio images similar to those observed. The parallel-case may produce
radio images unlike any observed SNR (center-brightened or with a
double-peak structure not describable as a shell). This is in agreement
with the findings of \citet{1990ApJ...357..591F}.

2. In models with gradients of the ambient density, the dependence of the
injection efficiency on the shock velocity (through the parameter $b$
defined in Sect. \ref{e-acc}) affects the degree of asymmetry of the radio
images: the asymmetry increases with increasing value of $b$.

3. Small variations of the ambient magnetic field lead to significant
asymmetries in the morphology of BSNRs (see Figs.~\ref{fig6} and
\ref{fig7}). Therefore, we conclude that the close similarity of the
radio brightness of the opposed limbs of a BSNR (i.e. $R\rs{max}\approx
1$ and $\theta\rs{D}\approx 180^o$) is evidence of uniform ambient \mag
field where the remnant expands.

4. Variations of the ambient density lead to asymmetries of the
remnant with extent comparable to that caused by non-uniform ambient
magnetic field if $b = 2$.

5. Strongly asymmetric BSNRs (i.e. $R\rs{max} \gg 1$ or $\theta\rs{D}
\ll 180^o$) imply either moderate variations of \mag or strong
(moderate) variations of the ISM density if $b < 2$ ($b\geq 2$) as in
the case, for instance, of interaction with a giant molecular cloud.

6. BSNRs with different intensities of the emission of the radio arcs
(i.e. $R\rs{max}>1$) can be produced by models with a gradient of
density or of magnetic field strength perpendicular to the arc (upper
panels in Fig.~\ref{fig6} and lower panels in Fig.~\ref{fig7}), and
the brightest arc is in the region of higher plasma density or higher
magnetic field strength.

7. Remnants with two slanting similar arcs (i.e. $\theta\rs{D} < 180^o$)
can be produced by models with a gradient of density or of magnetic
field strength running centered between the two arcs (lower panels in
Fig.~\ref{fig6} and upper panels in Fig.~\ref{fig7}), and the region
of convergence is where either the plasma density or the magnetic field
strength is higher.

8. In all the cases examined, the symmetry axis of the remnant is always
aligned with the gradient of density or of magnetic field.

We found that the degree of ordering of the magnetic field downstream
of the shock does not affect significantly the asymmetries induced
by gradients either of ambient plasma density or of ambient magnetic
field strength; thus our conclusions, derived in the case of disordered
magnetic field, do not change in the case of ordered magnetic field.

We defined useful model parameters to quantify the degree of asymmetry
of the remnants. These parameters may provide a powerful diagnostic in
the comparison between models and observations, as we have shown in a
few cases drawn from a randomly selected sample of BSNRs presented in
Table \ref{tab2}. For instance, if the density of the external medium
is known by other means (e.g. thermal X-rays, HI and Co maps, etc.),
BSNRs can be very useful to investigate the variation of the efficiency
of electron injection with the angle between the shock normal and the
ambient magnetic field or to investigate the dependence of the injection
efficiency from the shock velocity. Alternatively, BSNRs can be used as
probes to trace the local configuration of the galactic magnetic field
if the dependence of the injection efficiency from the obliquity is
known.

It is worth emphasizing that our model follows the evolution of the
remnant during the adiabatic phase and, therefore, their applicability
is limited to this evolutionary stage. In the radiative phase, the high
degree of compression suggested by radiative shocks leads to increase
of the radio brightness due to compression of ambient magnetic field
and electrons. Since our model neglects the radiative cooling it is
limited to relatively small compression ratios and, therefore, it is
not able to simulate this mechanism of limb brightening.

It will be interesting to expand the present study, considering the
detailed comparison of model results with observations. This may lead to a
major advance in the study of interactions between the magnetized ISM and
the whole galactic SNR population (not only BSNRs), since the mechanisms
at work in the BSNRs are also valid for SNRs of more complex morphology.

\bigskip
\acknowledgements{We thank the referee for constructive and helpful
criticism. The software used in this work was in part developed by the
DOE-supported ASC/Alliance Center for Astrophysical Thermonuclear Flashes
at the University of Chicago. The simulations have been executed at CINECA
(Bologna, Italy) in the framework of the INAF-CINECA agreement. This
work was supported by Ministero dell'Istruzione, dell'Universit\`a e
della Ricerca, by Istituto Nazionale di Astrofisica, and by Agenzia
Spaziale Italiana (ASI/INAF I/023/05/0).}

\appendix
\section{Derivation of Eq.~(\ref{K-evol})}
\label{app1}

We follow \cite{1998ApJ...493..375R} in the description of the evolution
of electron distribution. His approach is extended here to the possibility
to deal with non-uniform ISM (cf. \citealt{2006A&A...460..375P}). Fluid
element $a\equiv R(t\rs{i})$ was shocked at time $t\rs{i}$, where $R$
is the radius of the shock, and $a$ is the Lagrangian coordinate. At
that time the electron distribution on the shock was

\begin{equation}
 N(E\rs{i},a,t\rs{i})=K\rs{s}(a,t\rs{i})E\rs{i}^{-\zeta}~,
\end{equation}

\noindent
where $E\rs{i}$ is the electron energy at time $t\rs{i}$, $K\rs{s}$
is the normalization of the electron distribution immediately after the
shock (in the following, index ``s'' refers to the immediately post-shock
values), and $\zeta$ is the power law index. Since we are interested in
radio emission, we have to account for only energy losses of electrons
due to the adiabatic expansion (\citealt{1998ApJ...493..375R}):

\begin{equation}
 \frac{dE}{dt}=\frac{E}{3\rho}\frac{d\rho}{dt},
\end{equation}

\noindent
where $\rho$ is the mass density, so the energy varies as

\begin{equation}
 E=E\rs{i}\left(\frac{\rho(a,t)}{\rho\rs{s}(a,t\rs{i})}\right)^{1/3}.
\end{equation}

\noindent
The conservation law for the number of particles per unit volume per
unit energy interval

\begin{equation}
 N(E,a,t)=N(E\rs{i},a,t\rs{i}){a^2\,da\,dE\rs{i}\over \sigma
r^2\,dr\,dE},
\end{equation}

\noindent
where $\sigma$ is the shock compression ratio and $r$ is the
Eulerian coordinate, together with the continuity equation $\rho_{\rm
o}(a)a^2da=\rho(a,t)r^2dr$ (index ``o'' refers to the pre-shock values)
and the derivative

\begin{equation}
 \frac{dE\rs{i}}{dE} =
 \left(\frac{\rho(a,t)}{\rho\rs{s}(a,t\rs{i})}\right)^{-1/3},
\end{equation}

\noindent
implies that downstream

\begin{equation}
 N(E,a,t) = K\rs{s}(a,t\rs{i})E^{-\zeta}
 \left(\frac{\rho(a,t)}{\rho\rs{s}(a,t\rs{i})}\right)^{(\zeta+2)/3}~.
\end{equation}

\noindent
If $K_{\rm s}\propto \rho\rs{s} V\rs{sh}(t)^{-b}$, where $V\rs{sh}(t)$
is the shock velocity and $\rho\rs{s}$ is the immediately post-shock
value of density, then

\begin{equation}
 K\rs{s}(a,t\rs{i}) = K\rs{s}(R,t)
 \left(\frac{\rho\rs{o}(a)}{\rho\rs{o}(R)}\right)
 \left(\frac{V\rs{sh}(t)}{V\rs{sh}(t\rs{i})}\right)^{b}.
\end{equation}

\noindent
Therefore, the distribution of relativistic electrons follows

\begin{equation}
 \frac{K(a,t)}{K\rs{s}(R,t)}=\frac{N(E,a,t)}{N(E,R,t)}=
 \left(\frac{\rho\rs{o}(a)}{\rho\rs{o}(R)}\right)
 \left(\frac{V\rs{sh}(t)}{V\rs{sh}(t\rs{i})}\right)^{b}
 \left(\frac{\rho(a,t)}{\rho\rs{s}(a,t\rs{i})}\right)^{(\zeta+2)/3}.
 \label{KK}
\end{equation}

\noindent
Now we can substitute Eq. \ref{KK} with the ratio of the shock
velocities which comes from the expression (\citealt{1999A&A...344..295H})

\begin{equation}
 \frac{P(a,t)}{P\rs{s}(R,t)} =
 \left(\frac{\rho\rs{o}(a)}{\rho\rs{o}(R)}\right)^{-2/3}
 \left(\frac{V\rs{sh}(t\rs{i})}{V\rs{sh}(t)}\right)^2
 \left(\frac{\rho(a,t)}{\rho\rs{s}(R,t)}\right)^{5/3}~.
\end{equation}

\noindent
Thus, finally

\begin{equation}
 \frac{K(a,t)}{K\rs{s}(R,t)}=
 \left(\frac{P(a,t)}{P\rs{s}(R,t)}\right)^{-b/2}
 \left(\frac{\rho\rs{o}(a)}{\rho\rs{o}(R)}\right)^{-(b+\zeta-1)/3}
 \left(\frac{\rho(a,t)}{\rho\rs{s}(R,t)}\right)^{5b/6+(\zeta+2)/3}.
\end{equation}

\noindent
This formula may easily be used to calculate the profile of $K(a)$ for
known $P(a)$ and $\rho(a)$ in the case of the radial flow of fluid.  In
the case when mixing is allowed, the position $\vec R$ should correspond
to the same part of the shock which was at $\vec a$ at time $t\rs{i}$.

\bibliographystyle{aa}
\bibliography{references}

\end{document}